\overfullrule=0pt
\input harvmac
\def\a{{\alpha}}
\def\ah{{\widehat\alpha}}

\def\l{{\lambda}}
\def\lh{{\widehat\lambda}}

\def\b{{\beta}}
\def\bh{{\widehat\beta}}
\def\g{{\gamma}}
\def\gh{{\widehat\gamma}}
\def\d{{\delta}}
\def\dh{{\widehat\delta}}

\def\s{{\sigma}}
\def\r{{\rho}}
\def\sh{{\widehat\sigma}}
\def\rh{{\widehat\rho}}
\def\N{{\nabla}}
\def\Nb{{\overline\nabla}}
.
\def\O{{\Omega}}
\def\Ob{{\overline\O}}

\def\o{{\omega}}
\def\oh{{\widehat\omega}}

\def\ve{{\varepsilon}}
\def\veh{{\widehat\varepsilon}}

\def\p{{\partial}}
\def\pb{{\overline\partial}}
\def\t{{\theta}}
\def\th{{\widehat\theta}}
\def\ph{{\widehat p}}
\def\oh{{\widehat\o}}

\def\dhh{{\widehat d}}

\def\Pib{{\overline\Pi}}

\def\HH{{\cal H}}

\def\S{{\Sigma}}
\def\Sh{{\widehat\Sigma}}
\def\dhh{{\widehat d}}
\def\Ch{\widehat C}

\def\KK{{\cal K}}
\def\HH{{\cal H}}

\def\PP{{\cal P}}
\def\muh{{\widehat\mu}}

\baselineskip12pt

\Title{ \vbox{\baselineskip12pt
}}
{\vbox{\centerline{On-shell type II supergravity from the ambitwistor}
\bigskip
\centerline{pure spinor string}
}}
\smallskip
\centerline{Osvaldo Chandia\foot{e-mail: ochandiaq@gmail.com}, }
\smallskip
\centerline{\it Departamento de Ciencias, Facultad de Artes Liberales, Universidad Adolfo Ib\'a\~nez}
\centerline{\it Facultad de Ingenier\'{\i}a y Ciencias, Universidad Adolfo Ib\'a\~nez}
\centerline{\it Diagonal Las Torres 2640, Pe\~nalol\'en, Santiago, Chile}
\bigskip
\centerline{Brenno Carlini Vallilo\foot{e-mail: vallilo@gmail.com}, }
\smallskip
\centerline{\it Departamento de Ciencias F\'{\i}sicas, Facultad de Ciencias Exactas}
 \centerline{\it Universidad Andres Bello, Rep\'ublica 220, Santiago, Chile}

\bigskip
\bigskip
\noindent
We obtain all the type II supergravity constraints in the pure
spinor ambitwistor string by imposing consistency of local worldsheet
gauge symmetries.

\Date{November 2015}


\lref\MasonSVA{
  L.~Mason and D.~Skinner,
  ``Ambitwistor strings and the scattering equations,''
JHEP {\bf 1407}, 048 (2014).
[arXiv:1311.2564 [hep-th]].
}

\lref\CachazoGNA{
  F.~Cachazo, S.~He and E.~Y.~Yuan,
  ``Scattering equations and Kawai-Lewellen-Tye orthogonality,''
Phys.\ Rev.\ D {\bf 90}, no. 6, 065001 (2014).
[arXiv:1306.6575 [hep-th]].
}

\lref\CachazoHCA{
  F.~Cachazo, S.~He and E.~Y.~Yuan,
  ``Scattering of Massless Particles in Arbitrary Dimensions,''
Phys.\ Rev.\ Lett.\  {\bf 113}, no. 17, 171601 (2014).
[arXiv:1307.2199 [hep-th]].
}

\lref\CachazoIEA{
  F.~Cachazo, S.~He and E.~Y.~Yuan,
  ``Scattering of Massless Particles: Scalars, Gluons and Gravitons,''
JHEP {\bf 1407}, 033 (2014).
[arXiv:1309.0885 [hep-th]].
}

\lref\AdamoWEA{
  T.~Adamo, E.~Casali and D.~Skinner,
  ``A Worldsheet Theory for Supergravity,''
JHEP {\bf 1502}, 116 (2015).
[arXiv:1409.5656 [hep-th]].
}

\lref\ChandiaSFA{
  O.~Chandia and B.~C.~Vallilo,
  ``Ambitwistor pure spinor string in a type II supergravity background,''
JHEP {\bf 1506}, 206 (2015).
[arXiv:1505.05122 [hep-th]].
}

\lref\BerkovitsXBA{
  N.~Berkovits,
  ``Infinite Tension Limit of the Pure Spinor Superstring,''
JHEP {\bf 1403}, 017 (2014).
[arXiv:1311.4156 [hep-th], arXiv:1311.4156].
}

\lref\BerkovitsUE{
  N.~Berkovits and P.~S.~Howe,
  ``Ten-dimensional supergravity constraints from the pure spinor formalism for the superstring,''
Nucl.\ Phys.\ B {\bf 635}, 75 (2002).
[hep-th/0112160].
}

\lref\BerkPrivate{
  N.~Berkovits, private communication.
  }

\lref\ChandiaIX{
  O.~Chandia,
 ``A Note on the classical BRST symmetry of the pure spinor string in a curved background,''
JHEP {\bf 0607}, 019 (2006).
[hep-th/0604115].
}

\lref\JusinskasQJD{
  R.~L.~Jusinskas,
 ``Notes on the ambitwistor pure spinor string,''
JHEP {\bf 1605}, 116 (2016).
[arXiv:1604.02915 [hep-th]].
}

\newsec{Introduction}

In the last few years a new type of string theory has been
proposed \MasonSVA. In contrast with traditional string theories,
this new model contains only a finite number of states, namely, only
the massless modes. Also, it has been shown that this string theory
reproduces the scattering formul\ae\ of Cachazo {\it et.al.}
\refs{\CachazoGNA,\CachazoHCA,\CachazoIEA} for tree level massless field
theories in ten dimensions.

The state-operator correspondence of conformal field theories imply
that the states used to compute scattering amplitudes can also be used
to deform the worldsheet action. This was investigated in \AdamoWEA\
where it was found that the closure of worldsheet symmetry algebra
gives the bosonic part of the ten dimension supergravity equations of
motion. The most interesting aspect of the computation is that it is
an exact result. There are no $\alpha'$ corrections to these
equations. This comes from the fact that there are no massive states
to be integrated over.

In a previous paper \ChandiaSFA\ we described how to couple pure
spinor version of the ambitwistor string \BerkovitsXBA\ to a general
type II supergravity background. We found BRST symmetry was not enough
to impose constraints on the background, as opposed to the usual case
\BerkovitsUE. In flat space it appears that only BRST symmetry is
enough to impose on-shell conditions on the vertex operators. However,
we found out that there exist additional local symmetries which cannot
be ignored in curved space. We called the generators of these
symmetries $\cal K$ and $\cal H$. $\cal K$ is a ghost
number one charge and $\cal H$ is a conformal weight two
current. Nilpotency of the BRST charge $Q$ determined the nilpotency
constraints of \BerkovitsUE\ associated to the torsion and the
curvature. While the nilpotency constraints of \BerkovitsUE\
associated to the Kalb-Ramond field came from BRST invariance of
$\KK$. Then, nilpotency of $Q+ \KK$ should provide all the nilpotency
constraints of \BerkovitsUE. In this case, $Q+ \KK$ should be the BRST
charge of the theory \BerkPrivate. It remains to find a new $\HH$
which is BRST invariant under the new BRST charge.
We will construct such operator here and its BRST invariance gives
the holomorphicity constraints of \BerkovitsUE. Note that, the model the
we consider is holomorphic, then holomorphicity of the BRST current
is automatic. In \ChandiaSFA, it was suggested that the constraints associated
to the conservation of the BRST current in the normal pure spinor string
in a curved background \BerkovitsUE, are replaced by the existence of a BRST invariant
and conformal weight two world-sheet field, which corresponds to $\HH$. It would be interesting to understand the origin of this field and
to relate it with the mass-shell condition. In this way, we
complete the program suggested in \ChandiaSFA.

This paper is organized as follows. In Section 2 we give a brief
description of the ambitwistor pure spinor string in flat
background. We also describe the necessary modifications to
supersymmetry transformations in order to preserve the new BRST
charge. In Section 3 we discuss the supergravity background and the new
$\HH$ constraint. Then we show that nilpotency of the new BRST change
and BRST invariance of the $\HH$ imply all supergravity constraints. We
conclude the paper in Section 4 with some open questions to be
addressed in the future.

\newsec{ Ambitwistor superstring in a flat background }

The ambitwistor superstring in a flat background is described by the world-sheet action
\eqn\szero{ S = \int\! d^2z ~\left( P_m \pb X^m + p_\a \pb \t^\a + \ph_\ah \pb \th^\ah + \o_\a \pb \l^\a + \oh_\ah \pb \lh^\ah \right) ,}
where $(X,\t,\th)$ are the coordinates of the flat ambitwistor superspace, $(P,p,\ph)$ are the momentum conjugate variables of $(X,\t,\th)$. The pure spinor variables $(\l,\lh)$ are constrained by $\l\g^m\l=\lh\g^m\lh=0$, where the matrices $\g^m_{\a\b}$ and $\g^m_{\ah\bh}$ are the $16\times16$ symmetric gamma matrices in ten dimensions. The variables $(\o,\oh)$ are the momentum conjugate variables of the pure spinors. In \BerkovitsXBA, the usual pure spinor BRST charge was used to quantize this system. The BRST charge is given by
\eqn\qbflat{ Q = \oint \left( \l^\a d_\a + \lh^\ah \dhh_\ah \right) ,}
where
\eqn\dold{ d_\a = p_\a - \ha P_m (\g^m\t)_\a,\quad \dhh_\ah = \ph_\ah - \ha P_m (\g^m\th)_\ah .}

It was noted in  \ChandiaSFA\ that the action \szero\ has  a symmetry generated by
\eqn\newsym{ \oint \left( \l^\a (\p X^m B_{m\a} + \p\th^\bh B_{\bh\a} +\cdots) + \lh^\ah ( \p X^m B_{m\ah} + \p\t^\b B_{\b\ah} +\cdots) \right) ,}
where $B$ is the Kalb-Ramond potential in flat space and the terms $\cdots$ make \newsym\ supersymmetric and BRST invariant. Note that \szero\ has a further BRST invariant symmetry generated by
\eqn\Hzero{ \HH = -\ha P_m P^m .}

In \ChandiaSFA, the action \szero, the BRST charge \qbflat\ and the symmetry \newsym\ were generalized to curved background. The background supergeometry is constrained by imposing BRST invariance. It was found that nilpotency of the BRST charge and BRST invariance of the curved space version of \newsym\ impose the so-called `nilpotency' constraints of type II supergravity in \BerkovitsUE. Part of the remaining constraints of type II supergravity were determined in \ChandiaSFA\ by imposing BRST invariance of the curved space version of \Hzero. The `holomorphic' constraints of \BerkovitsUE\ involving $H=dB$ were not obtained. Our purpose is to determine these constraints. To achieve this, it is necessary to modify the BRST charge and supersymmetry in flat space first. The idea is that the BRST charge is the sum of \qbflat\ and the generator of \newsym. Another way to see this is to modify \dold\  to
\eqn\dnews{d_\a = p_\a - \ha (P_m - \p X_m) (\g^m\t)_\a + {1\over4} (\t\g_m\p\t) (\g^m\t)_\a ,}
$$ \dhh_\ah = \ph_\ah - \ha (P_m + \p X_m) (\g^m\th)_\ah - {1\over4} (\th\g_m\p\th) (\g^m\th)_\ah .$$
It is necessary that $(p,\ph, P)$ transform under space-time supersymmetry to make $(d,\dhh)$ supersymmetry invariant. The supersymmetry transformations are given by
\eqn\susy{ \d\t^\a =\ve^\a,\quad \d\th^\ah = \veh^\ah, \quad \d X^m = -\ha \ve\g^m\t-\ha \veh\g^m\th .}
The supersymmetric generalization of $\p X^m$ is
\eqn\Piflat{ \Pi^m = \p X^m + \ha \t\g^m\p\t + \ha \th\g^m\p\th ,}
and similarly for $\pb X^m$. It turns out that $d, \dhh$ and the action are supersymmetric if  $(p,\ph, P)$ transform as
\eqn\dSd{ \d p_\a = \ha (P_m - \p X_m) (\g^m\ve)_\a + {1\over4} (\ve\g^m\t) (\g_m\p\t)_\a ,}
\eqn\dSdh{ \d \ph_\ah = \ha (P_m + \p X_m) (\g^m\veh)_\ah - {1\over4} (\veh\g^m\th) (\g_m\p\th)_\ah ,}
\eqn\dSP{\d P_m = \ha \ve\g_m\p\t - \ha \veh\g_m\p\th .}
Besides $\Pi, d, \dhh$, there are other two supersymmetric combinations. They are
\eqn\Psusy{ P^-_m = P_m - \p X_m - \t\g_m\p\t,\quad P^+_m = P_m + \p X_m + \th\g_m \p\th .}
Note that the supersymmetric world-sheet fields $P^\pm, \Pi$ are not independent. They satisfy $P^+_m-P^-_m=2\Pi_m$.

The non zero OPE's between the supersymmetric variables are
\eqn\opes{ d_\a(y) d_\b(z) \to - {1\over(y-z)} \g^m_{\a\b} P^-_m,\quad \dhh_\ah(y) \dhh_\bh(z) \to - {1\over(y-z)} \g^m_{\ah\bh} P^+_m ,}
\eqn\opesa{ d_\a(y) \Pi^m (z) \to {1\over(y-z)} (\g^m\p\t)_\a,\quad \dhh_\a(y) \Pi^m (z) \to {1\over(y-z)} (\g^m\p\th)_\ah ,}
\eqn\opesb{d_\a(y) P^-_m(z) \to - {1\over(y-z)} 2 (\g_m\p\t)_\a,\quad \dhh_\ah(y) P^+_m(z) \to  {1\over(y-z)} 2 (\g_m\p\th)_\ah ,}
\eqn\opesc{ \Pi^m(y) P^-_n(z) \to - {1\over(y-z)^2} \d^m_n,\quad \Pi^m(y) P^+_n(z) \to - {1\over(y-z)^2} \d^m_n ,}
\eqn\opesd{ P^-_m(y) P^-_n(z) \to {1\over(y-z)^2} 2\eta_{mn},\quad P^+_m(y) P^+_n(z) \to -{1\over(y-z)^2} 2\eta_{mn} .}

Using the supersymmetric variables, the action \szero\ can be written as
\eqn\sone{ S = \int d^2z ~ [~ \ha (P^-_m+P^+_m)\Pib^m  + d_\a \pb\t^\a + \dhh_\ah \pb\th^\ah + \o_\a\pb\l^\a + \oh_\ah\pb\lh^\ah }
$$+ \ha(\t\g_m\p\t) \Pib^m - \ha (\t\g_m\pb\t) \Pi^m - \ha(\th\g_m\p\th) \Pib^m + \ha (\th\g_m\pb\th) \Pi^m $$
$$- {1\over4} (\t\g^m\p\t) (\th\g_m\pb\th) + {1\over4} (\t\g^m\pb\t) (\th\g_m\p\th) ~ ].$$
Note that the last two lines can be expressed as $\int \Pi^A \Pib^B B_{BA}$, where $B$ is the Kalb-Ramond field in a flat space-time. We now verify that the action \sone\ is supersymmetric and BRST invariant. Under the former, the action changes to
$$\int d^2z ~ [ ~ \ha (\ve\g_m\p\t) \Pib^m - \ha (\ve\g_m\pb\t) \Pi^m - \ha (\veh\g_m\p\th) \Pib^m + \ha (\veh\g_m\pb\th) \Pi^m $$
$$- {1\over4} (\ve\g^m\p\t) (\th\g_m\pb\th) - {1\over4} (\t\g^m\p\t) (\veh\g_m\pb\th) + {1\over4} (\ve\g^m\pb\t) (\th\g_m\p\th) + {1\over4} (\t\g^m\pb\t) (\veh\g_m\p\th) ~].$$
Consider the terms involving $\ve$. The terms with $X$ are proportional to
$$\int d^2z ~ \left( (\ve\g_m\p\t) \pb X^m - (\ve\g_m \pb\t) \p X^m  \right),$$
which vanishes after integrating by parts. The other terms involving $\ve$ are proportional to
$$\int d^2z ~ \left( (\ve\g_m\p\t) (\t\g^m\pb\t) - (\ve\g_m\pb\t) (\t\g^m\p\t) \right).$$
If we use the Fierz identity for gamma matrices\foot{Namely, $\g^m_{(\a\b}\g^n_{\g)\d}\eta_{mn}=0$.}, this expression is equal to
$$\int d^2z ~ (\ve\g_m\t) (\p\t\g^m\pb\t) .$$
If, instead of this, we integrate by parts we obtain that the same expression is equal to
$$\int d^2z ~ -2  (\ve\g_m\t) (\p\t\g^m\pb\t) .$$
Therefore, something that is equal to $-2$ times itself has to vanish. A similar calculation is obtained for the terms involving $\veh$ in the supersymmetric variation of \sone.

To verify the BRST variation of the action, we need the BRST transformations of the world-sheet fields. They can be obtained by using the OPE's \opes-\opesd\foot{The BRST transformation of a world-sheet field $\Psi$ is defined and computed as  $Q\Psi (z)\equiv\oint dy ~j(y)\Psi(z)$, where $j$ is the integrand of the BRST charge \qbflat\ with the definitions \dnews.}  . We obtain that $(\l,\lh)$ are invariant, $Q\t=\l, Q\th=\lh, Q\o=d, Q\oh=\dhh$ and
\eqn\Qwsflat{Q\Pi^m = \l\g^m\p\t+\lh\g^m\p\th,\quad QP^-_m=-2\l\g_m\p\t,\quad QP^+_m=2\lh\g_m\p\th ,}
$$ Q d_\a = -(\l\g^m)_\a P^-_m,\quad Q \dhh_\ah = -(\lh\g^m)_\ah P^+_m .$$
Using the above transformations it is straightforward to verify that the action is invariant under the BRST transformations. Note that the variable $P_m$ is no longer BRST invariant. Its transformation is given by
\eqn\QBP{Q P_m = - \ha \p (\l\g_m\t - \lh \g_m \th ) .}

The immediate consequence is that \Hzero\ is no longer BRST invariant and needs to be modified. It is replaced by
\eqn\newH{ \HH = - {1\over4}  P^-_m  P^m_- - {1\over4}  P^+_m  P^m_+  + d_\a \p\t^\a - \dhh_\ah \p\th^\ah + \o_\a\p\l^\a - \oh_\ah\p\lh^\ah ,}
which is supersymmetric and BRST invariant. Note that $\HH$ looks like the stress-tensor but it is independent of it.

In the next section we will study the above model in a curved background. The type II supergravity constraints will be obtained by the nilpotency of the BRST charge and BRST invariance of the curved space version of \newH.

\newsec{ Ambitwistor superstring on a type II supergravity background}

In this section we study the model described above in a generic curved background. The action is the covariantization of \sone, that is
\eqn\SC{ S = \int d^2z ~ \left( \PP_a \Pib^a - \Pi^A \Pib^B B_{BA} + d_\a \Pib^\a + \dhh_\ah \Pib^\ah + \o_\a \Nb \l^\a + \oh_\ah \Nb \lh^\ah \right) ,}
where $\Pib^A = \pb Z^M E_M{}^A$, with $E_M{}^A$ being the vielbein superfield, and $Z^M$ are the coordinates of the curved ten-dimensional superspace. The covariant derivatives are defined with the background Lorentz connections, that is,
\eqn\cov{\Nb \l^\a = \pb \l^\a + \l^\b \Ob_\b{}^\a,\quad \Nb \lh^\ah = \pb \lh^\ah + \lh^\bh \Ob_\bh{}^\ah ,}
where $\Ob_\b{}^\a = \pb Z^M \O_{M\b}{}^\a$ and $\Ob_\bh{}^\ah = \pb Z^M \O_{M\bh}{}^\ah$.

Note that \SC\ is reduced to \sone\ in the flat space limit. To verify
this, we express $\PP, d, \dhh$ in terms of canonical conjugate
variables $(\l, \o), (\lh, \oh)$ and $(Z^M, P_M)$. Recall that $P_M$ is given  by
$$
(-1)^M P_M = E_M{}^a \PP_a - E_M{}^\a d_\a - E_M{}^\ah \dhh_\a + \Pi^A B_{AM} + \O_{M\a}{}^\b \l^\a\o_\b+\O_{M\ah}{}^\bh\lh^\ah\oh_\bh .
$$
The values of the background fields in flat space are known. The spin connection $\O$ vanishes, the non-zero components of the $B$ field are
$$
B_{m\a}=\ha(\g_m\t)_\a,\quad B_{m\ah}=-\ha(\g_m\th)_\ah,\quad B_{\a\bh}=-{1\over4}(\g^m\t)_\a(\g_m\th)_\bh ,$$
and the non-zero components of the vielbein are
$$
E_m{}^a=\d_m^a,\quad E_\mu{}^\a=\d_\mu{}^\a,\quad E_\muh{}^\ah=\d_\muh^\ah,\quad E_\mu{}^a=-\ha(\g^a\t)_\mu,\quad E_\muh{}^a=-\ha(\g^a\th)_\muh .$$
Using these values, one can check that $d, \dhh$ become \dnews, and $\PP$ becomes
$\ha(P_- + P_+)$.

A field redefinition of $d$, $\dhh$ and $\PP$ can
eliminate $B_{AB}$ from the action, but this changes $Q$ and $\HH$
defined bellow. We found it is simpler to keep $B_{AB}$ in \SC.

The BRST charge is
\eqn\QBC{ Q = \oint \left( \l^\a d_\a + \lh^\a \dhh_\ah \right) .}
The BRST transformations\foot{The BRST transformation for a world-sheet field $\Psi$ is defined and obtained as $Q\Psi=\oint \{j , \Psi \}$ where $\{\cdots\}$ represents the Poisson bracket which is canonical for the pairs $(P,Z), (\l.\oh)$ and $(\lh,\oh)$. Some detail about these kind of computations can be found in \ChandiaIX.  }
of the pure spinor variables are
\eqn\Qpurel{ Q \l^\a = - \l^\b \S_\b{}^\a,\quad Q \o_\a = d_\a + \S_\a{}^\b \o_\b ,}
\eqn\Qpurelh{ Q \lh^\ah = - \lh^\bh \Sh_\bh{}^\ah,\quad Q \oh_\ah = d_\a + \Sh_\ah{}^\bh \oh_\bh ,}
where
\eqn\Lor{ \S_\a{}^\b = \l^\g \O_{\g\a}{}^\b + \lh^\gh \O_{\gh\a}{}^\b,\quad \Sh_\ah{}^\bh = \l^\g \O_{\g\ah}{}^\bh + \lh^\gh \O_{\gh\ah}{}^\bh,}
are field-dependent Lorentz rotations.

The BRST transformations of the other fields are given by
\eqn\Qd{ Q d_\a = ( \l^\b T_{\b\a}{}^a + \lh^\bh T_{\bh\a}{}^a ) \PP_a - (\l^\b T_{\b\a}{}^\g + \lh^\bh T_{\bh\a}{}^\g ) d_\g - ( \l^\b T_{\b\a}{}^\gh + \lh^\bh T_{\bh\a}{}^\gh ) \dhh_\gh }
$$ + ( \l^\b R_{\b\a\g}{}^\d + \lh^\bh R_{\bh\a\g}{}^\d ) \l^\g \o_\d + ( \l^\b R_{\b\a\gh}{}^\dh + \lh^\bh R_{\bh\a\gh}{}^\dh ) \lh^\gh \oh_\dh - \l^\b \Pi^A H_{A\b\a} - \lh^\bh \Pi^A H_{A\bh\a} + \S_\a{}^\b d_\b, $$
\eqn\Qdh{Q \dhh_\ah = ( \l^\b T_{\b\ah}{}^a + \lh^\bh T_{\bh\ah}{}^a ) \PP_a - (\l^\b T_{\b\ah}{}^\g + \lh^\bh T_{\bh\ah}{}^\g ) d_\g - ( \l^\b T_{\b\ah}{}^\gh + \lh^\bh T_{\bh\ah}{}^\gh ) \dhh_\gh }
$$ + ( \l^\b R_{\b\ah\g}{}^\d + \lh^\bh R_{\bh\ah\g}{}^\d ) \l^\g \o_\d + ( \l^\b R_{\b\ah\gh}{}^\dh + \lh^\bh R_{\bh\ah\gh}{}^\dh ) \lh^\gh \oh_\dh - \l^\b \Pi^A H_{A\b\ah} - \lh^\bh \Pi^A H_{A\bh\ah} + \S_\ah{}^\bh \dhh_\bh, $$
\eqn\QP{Q \PP_a = -( \l^\b T_{\b a}{}^b + \lh^\bh T_{\bh a}{}^b ) \PP_b + (\l^\b T_{\b a}{}^\g + \lh^\bh T_{\bh a}{}^\g ) d_\g + ( \l^\b T_{\b a}{}^\gh + \lh^\bh T_{\bh a}{}^\gh ) \dhh_\gh }
$$ - ( \l^\b R_{\b a\g}{}^\d + \lh^\bh R_{\bh a\g}{}^\d ) \l^\g \o_\d - ( \l^\b R_{\b a\gh}{}^\dh + \lh^\bh R_{\bh a\gh}{}^\dh ) \lh^\gh \oh_\dh + \l^\b \Pi^A H_{A\b a} + \lh^\bh \Pi^A H_{A\bh a} + \S_a{}^b \PP_b .$$
where $T^A$ is the torsion superfield, $R_A{}^B$ is the curvature superfield and $H=dB$. Recall that the torsion and curvature super two-forms are given in terms of the super one-forms vielbein $E^A$ and connection $\O_A{}^B$ according to
$$
T^A = dE^A + E^B \O_B{}^A,\quad R_A{}^B = d\O_A{}^B+\O_A{}^C\O_C{}^B ,$$
where the wedge product between super forms is assumed and $d=dZ^M\p_M$ is the exterior derivative in superspace.

The computation of $Q^2$ can be obtained by applying $Q$ to itself and using \Qd\ and \Qdh. We obtain
\eqn\QQ{ Q^2 = \oint [~ \l^\a \l^\b ( T_{\a\b}{}^a \PP_a - T_{\a\b}{}^\g d_\g - T_{\a\b}{}^\gh \dhh_\gh + R_{\a\b\g}{}^\d \l^\g \o_\d + R_{\a\b\gh}{}^\dh \lh^\gh \oh_\dh - \Pi^A H_{A\a\b} ) }
$$+ \lh^\ah \lh^\bh ( T_{\ah\bh}{}^a \PP_a - T_{\ah\bh}{}^\g d_\g - T_{\ah\bh}{}^\gh \dhh_\gh + R_{\ah\bh\g}{}^\d \l^\g \o_\d + R_{\ah\bh\gh}{}^\dh \lh^\gh \oh_\dh - \Pi^A H_{A\ah\bh} )$$
$$ +2\l^\a \lh^\bh ( T_{\a\bh}{}^a \PP_a - T_{\a\bh}{}^\g d_\g - T_{\a\bh}{}^\gh \dhh_\gh + R_{\a\bh\g}{}^\d \l^\g \o_\d + R_{\a\bh\gh}{}^\dh \lh^\gh \oh_\dh - \Pi^A H_{A\a\bh} ) ~ ].$$
Therefore, the nilpotency of  $Q$ implies the constraints
\eqn\nilpa{\l^\a \l^\b T_{\a\b}{}^A = \l^\a \l^\b H_{\a\b A} = \l^\a \l^\b R_{\a\b\gh}{}^\dh = \l^\a \l^\b \l^\g R_{\a\b\g}{}^\d = 0,}
\eqn\nilpb{\lh^\ah \lh^\bh T_{\ah\bh}{}^A = \lh^\ah \lh^\bh H_{\ah\bh A} = \lh^\ah \lh^\bh R_{\ah\bh\g}{}^\d = \lh^\ah \lh^\bh \lh^\gh R_{\ah\bh\gh}{}^\dh = 0,}
\eqn\nilpc{T_{\a\bh}{}^A = H_{\a\bh A} = \l^\a \l^\b R_{\gh\a\b}{}^\d = \lh^\ah \lh^\bh R_{\g\ah\bh}{}^\dh = 0 ,}
which are part of the type II supergravity constraints.

The remaining type II supergravity constraints come from the BRST invariance of the curved space generalization of \newH. It is given by
\eqn\newH{ \HH =   -\ha \eta^{ab} \PP_a \PP_b + d_\a \dhh_\bh P^{\a\bh}+ d_\a \lh^\bh \oh_\gh \Ch_\bh{}^{\gh\a} + \dhh_\ah \l^\b \o_\g C_\b{}^{\g\ah} + \l^\a \o_\b \lh^\gh \oh_\dh S_{\a\gh}{}^{\b\dh},}
$$ + d_\a \Pi^\a - \dhh_\ah \Pi^\ah + \o_\a \N\l^\a - \oh_\ah \N \lh^\a - \ha \Pi_a \Pi^a .$$
The superfields $(P,C,\Ch,S)$ are constrained as consequence of the BRST invariance of $\HH$. It turns out that they are the same superfields appearing in \BerkovitsUE, that is, $P$ contains the Ramond-Ramond field-strengths of type II supergravity, $C,\Ch$ contain the dilatini and gravitini field-strengths of type II supergravity and $S$ contains the curvature of type II supergravity.

Using the BRST transformations of the world-sheet fields, one obtains that $Q$ acting on \newH\ gives
\eqn\QHH{ Q \HH = \ha \l^\a ( \PP^a \PP^b - \Pi^a \Pi^b )  T_{\a(ab)} + \ha \lh^\ah ( \PP^a \PP^b - \Pi^a \Pi^b ) T_{\ah(ab)} }
$$- \l^\a \PP^a \Pi^b H_{\a ab} - \lh^\ah \PP^a \Pi^b H_{\ah ab} + \l^\a ( \PP^a + \Pi^a ) d_\b  T_{a\a}{}^\b + \lh^\ah ( \PP^a - \Pi^a )  \dhh_\bh T_{a\ah}{}^\bh $$
$$- \l^\a ( \PP^a + \Pi^a ) \l^\b \o_\g R_{a\a\b}{}^\g - \l^\ah ( \PP^a - \Pi^a ) \lh^\bh \oh_\gh R_{a\ah\bh}{}^\gh $$
$$+ \l^\a ( ( \PP^a + \Pi^a ) \Pi^\b + \dhh_\gh \Pi^a P^{\b\gh} + \lh^\gh \oh_\dh \Pi^a \Ch_\gh{}^{\dh\b} ) ( T_{\b\a a} - H_{\b\a a} )  $$
$$- \lh^\ah ( ( \PP^a - \Pi^a ) \Pi^\bh - d_\g \Pi^a P^{\g\bh} + \l^\g \o_\d \Pi^a C_\g{}^{\d\bh} ) ( T_{\bh\ah a}+ H_{\bh\ah a} ) $$
$$+ 2 d_\b \lh^\dh ( \Pi^\rh + \ha d_\g P^{\g\rh}- \ha \l^\g \o_\s C_\g{}^{\s\rh} ) ( T_{\dh\rh}{}^\b + \ha P^{\b\ah}H_{\ah\dh\rh}) $$
$$- 2 \dhh_\bh \l^\d ( \Pi^\r + \ha \dhh_\gh P^{\r\gh}+ \ha \lh^\gh \oh_\sh \Ch_\gh{}^{\sh\r}  ) ( T_{\d\r}{}^\bh + \ha P^{\a\bh}H_{\a\d\r}) $$
$$+ \lh^\ah ( \PP^a + \Pi^a )  d_\b ( T_{a\ah}{}^\b - T_{\ah\gh a} P^{\b\gh} )  + \l^\a ( \PP^a - \Pi^a ) \dhh_\bh ( T_{a\a}{}^\bh + T_{\a\g a} P^{\g\bh} ) $$
$$+ \l^\ah ( \PP^a + \Pi^a ) \l^\b \o_\g  ( -R_{a\ah\b}{}^\g + T_{\ah\dh a} C_\b{}^{\g\ah} ) + \l^\a ( \PP^a - \Pi^a ) \lh^\bh \oh_\gh ( -R_{a\a\bh}{}^\gh + T_{\a\d a} \Ch_\bh{}^{\gh\d} ) $$
$$- 2 \lh^\dh \l^\b \o_\g ( \Pi^\rh + \ha d_\s P^{\s\rh} ) ( R_{\rh\dh\b}{}^\g - \ha C_\b{}^{\g\ah} H_{\ah\rh\dh} ) + 2 \l^\d \lh^\bh \oh_\gh ( \Pi^\r + \ha \dhh_\sh P^{\r\sh} ) ( R_{\r\d\bh}{}^\gh + \ha \Ch_\bh{}^{\gh\a} H_{\a\r\d} ) $$
$$+ \l^\a d_\b \dhh_\gh ( \N_\a P^{\b\gh} - T_{\a\d}{}^\b P^{\d\gh} + C_\a{}^{\b\gh} ) + \lh^\ah d_\b \dhh_\gh ( \N_\ah P^{\b\gh} - T_{\ah\dh}{}^\gh P^{\b\dh} - \Ch_\ah{}^{\gh\b} ) $$
$$- \lh^\ah d_\b \lh^\gh \oh_\dh ( \N_\ah \Ch_\gh{}^{\dh\b} + R_{\rh\ah\gh}{}^\dh P^{\b\rh} ) - \l^\a \dhh_\bh \l^\g \o_\d ( \N_\a C_\g{}^{\d\bh} - R_{\r\a\g}{}^\d P^{\r\bh} ) $$
$$+ \l^\a d_\b \lh^\gh \oh_\dh ( - R_{\rh\a\gh}{}^\dh P^{\b\rh} + T_{\a\r}{}^\b \Ch_\gh{}^{\dh\r} - \N_\a \Ch_\gh{}^{\dh\b} + S_{\a\gh}{}^{\b\dh} ) $$
$$+ \lh^\ah \dhh_\bh \l^\g \o_\d ( R_{\r\ah\g}{}^\d P^{\r\bh} + T_{\ah\rh}{}^\bh C_\g{}^{\d\rh} - \N_\ah C_\g{}^{\d\bh} + S_{\g\ah}{}^{\d\bh} ) $$
$$+ \l^\a \l^\b \o_\g \lh^\dh \oh_\rh ( R_{\s\a\b}{}^\g \Ch_\dh{}^{\rh\s} + R_{\sh\a\dh}{}^\rh C_\b{}^{\g\sh} + \N_\a S_{\b\dh}{}^{\g\rh} ) $$ $$+ \lh^\ah \l^\b \o_\g \lh^\dh \oh_\rh ( R_{\s\ah\b}{}^\g \Ch_\dh{}^{\rh\s} + R_{\sh\ah\dh}{}^\rh C_\b{}^{\g\sh} + \N_\ah S_{\b\dh}{}^{\g\rh} ) .$$
Therefore, BRST invariance of $\HH$ implies the holomorphic constraints for type II supergravity of \BerkovitsUE. They are,
\eqn\qjone{ T_{\a(ab)} = T_{a\a}{}^\b = T_{a\a}{}^\bh + T_{\a\g a} P^{\g\bh} = \l^\a \l^\b R_{a\a\b}{}^\g = 0 ,}
\eqn\qjHone{ H_{\a ab} = H_{\ah ab}= T_{\b\a a} - H_{\b\a a} = T_{\bh\ah a} + H_{\bh\ah a} = 0 ,}
\eqn\qjtwo{ T_{\ah(ab)} = T_{a\ah}{}^\bh = T_{a\ah}{}^\b - T_{\ah\gh a} P^{\b\gh} = \lh^\ah \lh^\bh R_{a\ah\bh}{}^\gh = 0 ,}
\eqn\qjthree{ R_{a\ah\b}{}^\g - T_{\ah\dh a} C_\b{}^{\g\ah} = R_{a\a\bh}{}^\gh - T_{\a\d a} \Ch_\bh{}^{\gh\d} = 0 ,}
\eqn\bhforHone{ T_{\g\a}{}^\bh + \ha P^{\d\bh} H_{\d\g\a} = T_{\gh\ah}{}^\b + \ha P^{\b\dh} H_{\gh\dh\ah} = 0 ,}
\eqn\bhforHtwo{ R_{\d\a\bh}{}^\gh + \ha \Ch_\bh{}^{\gh\r} H_{\r\d\a} = R_{\dh\ah\b}{}^\g - \ha C_\b{}^{\g\rh} H_{\rh\dh\ah} = 0.}
\eqn\qjfour{\N_\a P^{\b\gh} - T_{\a\d}{}^\b P^{\d\gh} + C_\a{}^{\b\gh} = \N_\ah P^{\b\gh} - T_{\ah\dh}{}^\gh P^{\b\dh} - \Ch_\ah{}^{\gh\b} = 0 ,}
\eqn\qjfive{ \N_\a \Ch_\gh{}^{\dh\b} - T_{\a\r}{}^\b \Ch_\gh{}^{\dh\r} + R_{\rh\a\gh}{}^\dh P^{\b\rh} - S_{\a\gh}{}^{\b\dh} = 0,}
\eqn\qjsix{ \N_\ah C_\g{}^{\d\bh} - T_{\ah\rh}{}^\bh C_\g{}^{\d\rh} - R_{\r\ah\g}{}^\d P^{\r\bh} - S_{\g\ah}{}^{\d\bh} = 0 ,}
\eqn\qjseven{ \l^\a \l^\b ( \N_\a C_\b{}^{\g\dh} - R_{\r\a\b}{}^\g P^{\r\dh} ) = \lh^\ah \lh^\bh ( \N_\ah \Ch_\bh{}^{\d\gh} + R_{\rh\ah\bh}{}^\gh P^{\d\rh} ) = 0 ,}
\eqn\qjeight{ \l^\a \l^\b  ( R_{\s\a\b}{}^\g \Ch_\dh{}^{\rh\s} + R_{\sh\a\dh}{}^\rh C_\b{}^{\g\sh} + \N_\a S_{\b\dh}{}^{\g\rh} ) = 0 ,}
\eqn\qjnine{ \lh^\ah \lh^\gh ( R_{\s\ah\b}{}^\g \Ch_\dh{}^{\rh\s} + R_{\sh\ah\dh}{}^\rh C_\b{}^{\g\sh} + \N_\ah S_{\b\dh}{}^{\g\rh} ) = 0 .}

The constraints \nilpa-\nilpc\ and \qjone-\qjnine\ imply that the background satisfies the equations of type II supergravity in ten-dimensional superspace \BerkovitsUE.

\newsec{Conclusion and further directions}
In this paper we have described how to obtain on-shell type II
supergravity from a reduced set of worldsheet constraints. The
computation is done at semi-classical level and it is likely that a
full quantum computation will require a modified version of the
constraints, as in \AdamoWEA. It would be very interesting to
understand how possible quantum corrections cancel in order to
preserve all the constraints. We plan to investigate this problem in
the future.

However, the most important issue to understand regards physical states and
vertex operators. The vertices discussed by Berkovits in
\BerkovitsXBA\ are no longer physical. Furthermore, the integrated
version does not appear to have $B_{mn}$ potential and its BRST
invariance does not fix completely the prepotentials $A_\alpha$ and
$\widehat A_{\widehat\alpha}$. If we compare with the usual pure spinor
string, the set of constraints we obtain from $\HH$ is the same set
obtained from conservation of the BRST charge \BerkovitsUE. This is
directly related to the action and we know that in the usual case
fluctuations of the action define the integrated vertex operator. It
is likely that for the ambitwistor string the integrated vertex
operator is given by fluctuations of $\HH$. This seems to be the case
for the original Mason-Skinner string. The unusual delta function in the integrated vertex operators seem to be related to the Lagrange multiplier
for the $\HH$ constraint. It would be useful to make this connection clearer in the path integral with the inclusion of a Nakanishi-Lautrup field.

Another interesting question is how the $AdS_5\times S^5$ case is
modified. In \ChandiaSFA\ we have shown that the sigma model for this
background is much simpler than the usual case, the global
$PSU(2,2|4)$ symmetry is promoted to a chiral Kac-Moody
symmetry. Furthermore, the $\HH$ constraint was shown to be the square
of the Kac-Moody current, like a Sugawara current. It is not clear if
this will continue to hold after the modifications discussed in the
present work. Since modified the supersymmetry transformation, it is
likely that we will have to modify the $PSU(2,2|4)$ symmetry.

{\bf Note added:} After the present work was made public, the paper \JusinskasQJD\ appered. In that work the author gives further evidence that the coholomology of the original BRST charge of \BerkovitsXBA\ does not give the correct spectrum. Following the ideas presented here, the author constructs a pair of $b$-ghosts and shows that the stress-energy tensor of the model described here is BRST-exact. We hope that we can use these results together with the ideas described in the paragraph above to construct the correct vertex operators for the ambitwistor pure spinor string.

\bigskip

\noindent
{\bf Acknowledgements:} We would like to thank Nathan Berkovits for
useful discussions and comments. This work is partially supported
by FONDECYT grants 1120263 and 1151409 and CONICYT grant DPI2014115.

\listrefs

\end